\documentclass[epj]{svjour}
\usepackage{graphics}
\begin{document}
\title{Random Delayed-Choice Quantum Eraser via Two-Photon Imaging}

\author{Giuliano Scarcelli\inst{1,2}\thanks{\emph{Present address:} Harvard Medical School and Wellman Center for
Photomedicine, Massachusetts General Hospital, Boston Ma 02114,
gscarcelli@partners.org} \and Yu Zhou \inst{1} \and Yanhua Shih \inst{1} 
%
}                     
%
%
\institute{Department of Physics, University of Maryland,
Baltimore County, Baltimore, Maryland 21250 \and Dipartimento
Interateneo di Fisica, Universita' di Bari, 70126, Bari, Italy}
\date{Received: date / Revised version: date}
%
\abstract{ We report on a delayed-choice quantum eraser experiment
based on a two-photon imaging scheme using entangled photon pairs.
After the detection of a photon which passed through a
double-slit, a random delayed choice is made to erase or not erase
the which-path information by the measurement of its distant
entangled twin; the particle-like and wave-like behavior of the
photon are then recorded simultaneously and respectively by only
one set of joint detection devices. The present eraser takes
advantage of two-photon imaging. The complete which-path
information of a photon is transferred to its distant entangled
twin through a ``ghost" image. The choice is made on the Fourier
transform plane of the ghost image between reading ``complete
information" or ``partial information" of the double-path.
\PACS{
      {PACS-key}{discribing text of that key}   \and
      {PACS-key}{discribing text of that key}
     } 
} 
\maketitle
\section{Introduction}
\label{intro} Quantum erasure was proposed in $1982$ by Scully and
Druhl \cite{scully}. After two decades the subject has become one
of the most intriguing topics in probing the foundations of
quantum mechanics \cite{ahoronov,amscie}. The idea of quantum
erasure lies in its connection to Bohr's principle of
complementarity \cite{bohr}: although a quantum mechanical object
is dually particle and wave; its particle-like and wave-like
behaviors cannot be observed simultaneously.  For example, if one
observes an interference pattern from a standard Young's
double-slit interferometer by means of single-photon counting
measurement, a photon must have been passing both slits like a
wave and consequently the which-slit information can never be
learned.   On the other hand, any information about through which
slit the photon has passed destroys the interference. In this
context Scully and Druhl showed that if the which-slit
(which-path) information is erased, the interference pattern can
be recovered; the situation becomes extremely fascinating when the
erasing idea is combined with the delayed choice proposal by
Wheeler and Alley \cite{wheeler,Alley}: i.e. even after the
detection of the quantum itself, it is still possible to decide
whether to erase or not to erase the which-path information, hence
to observe the wave behavior or the particle behavior of the
quantum mechanical object.

In the past two decades, a number of experiments demonstrated the
quantum eraser idea by means of different experimental approaches
and/or different point of theoretical concerns
\cite{chi,herzog,rempe,dopfer,schw,yoonho,tse,haroche,trifonov,bras,leuchs};
in particular Kim et al. \cite{yoonho} have realized an experiment
very close to the original proposal by using entangled photon pair
of spontaneous parametric down-conversion (SPDC). The experiment
demonstrated that the which-path information of a photon passing
through a double-slit can be erased at-a-distance by its entangled
twin even after the annihilation of the photon itself.   The
choice was made between the joint detection of a single two-photon
amplitude that involved either the upper slit or the lower slit
(read which-path information) or the joint detection of a pair of
indistinguishable two-photon amplitudes involving both slits
(erase which-path information).

Unlike all previous experiments the present work takes advantage
of two-photon imaging.  A photon passes through a standard Young's
double-slit for its complementarity examination. The quantum
correlation between this photon and its entangled twin allows the
formation of a ``ghost" image of the double-slit on the side of
the entangled twin. Thus, the which path information is completely
passed to the entangled twin photon and can be erased by the
detection of the twin.  After the detection of the photon which
passed through the double-slit, a random choice is made on the
Fourier transform plane of the ``ghost" image between ``reading
complete information" or ``reading partial information" of the
double path. Inherently, this new approach can be easily extended
to multiple paths or arbitrary continuous spatial modulations.

Any attempt to interpret the physics of the quantum eraser in
terms of complementarity examination on a single photon leads to
counterintuitive results and paradoxical conclusions; on the other
hand, if the two-photon nature of the phenomenon is accepted, a
straightforward explanation of the observed effect can be given
through Klyshko's interpretation of two-photon geometric optics.
From a new angle it is emphasized that the physics behind
two-photon phenomena is significantly different from that of two
independent photons\cite{shih}.

In this context another novelty of our experiment is particularly
important: a new type of detection scheme. In all previous quantum
erasers, the observation or not observation of the interference
pattern were associated to different experiments, or at least to
different photoelectric detectors. Therefore even though the
physics behind the erasure has been exploited, the implementation
of the random delayed choice can still be improved. In our
realization both the erasing choice and the reading choice are
analyzed by a single detector. This characteristic stresses the
interpretative difficulties of the quantum eraser. In fact in our
experiment the particle-like and wave-like behavior of the photon
are recorded randomly and simultaneously by the same pair of joint
measurement devices in only one measurement process.

The nearly equivalent experimental conditions in which the
realization of the different choices occurs fully implement
Wheeler and Alley delayed-choice proposal and therefore raise
troubling questions on where the measurement, hence the collapse
of the wave function, occurs. The validity of Bohr's principle of
complementarity is probed in a deeper way than in previous
experiments.

The paper is organized as follows: in section II, the principle
behind our quantum eraser is explained and its connection to the
intriguing physics of quantum imaging; in section III, a
mathematical derivation resulting in the experimental observation
is provided with a brief discussion of attempted interpretations
with their difficulties; in section IV, the experimental setup and
results are described in detail; and finally in section V, some
conclusive remarks are presented.

\section{Quantum imaging}

The quantum eraser here reported uses the fascinating physics of
quantum imaging.  The study of quantum imaging started ten years
ago after the first demonstration of an imaging experiment that
used entangled two-photon state of spontaneous parametric
down-conversion \cite{pittman,klyshko}.   In that experiment, the
signal photon of SPDC passed through an imaging lens and a
complicated aperture, while the idler photon propagated freely;
nevertheless the complete spatial distribution information was
present in the idler photon side of the setup and an image (named
``ghost" because even though it was formed by the idler radiation,
it reconstructed the spatial modulation experienced only by the
signal radiation) was formed in a plane satisfying a Gaussian thin
lens equation involving both arms of the setup. Over the past ten
years, quantum imaging has attracted a great deal of attention.
The equivalence between two-photon Fourier optics and classical
Fourier optics, (with the replacement of the two-photon amplitudes
leading to a joint detection by the spatial modes of the classical
electric field) has been shown \cite{sergienko}. The two-photon
amplitudes and their coherent superposition are troubling concepts
in a classical sense because they imply a non-local behavior of
the radiation; however, they explain in an elegant, consistent and
intuitive way all the features of entangled two-photon optics.

\begin{figure*}
\resizebox{0.75\textwidth}{!}{%
  \includegraphics{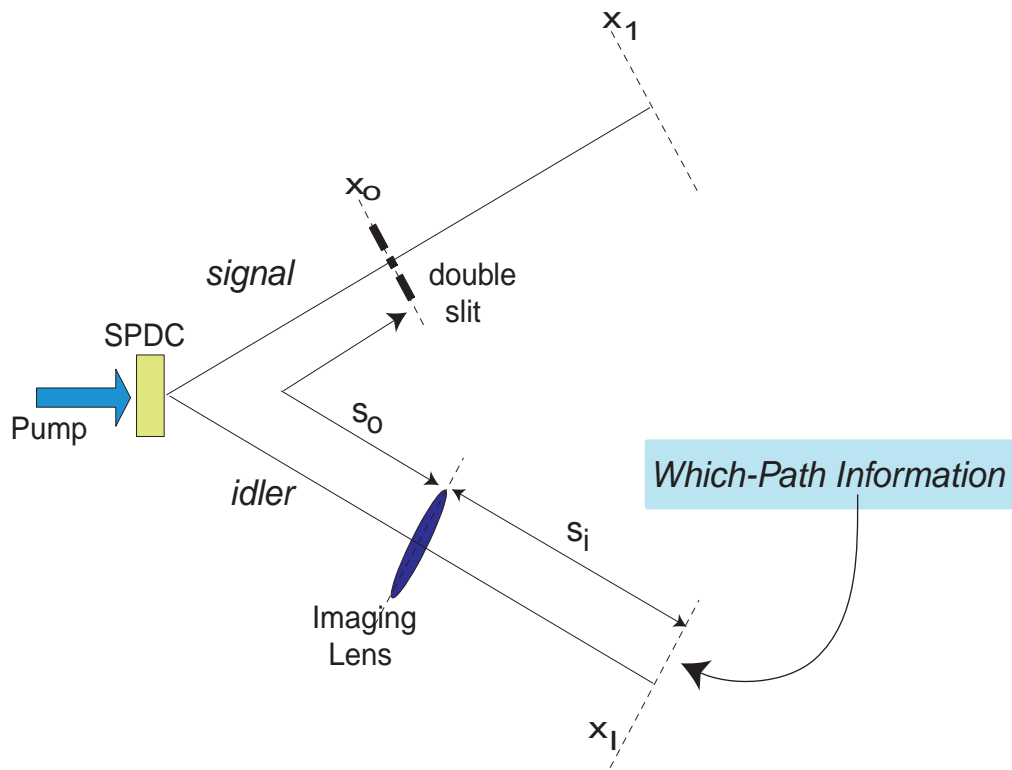}
}

\caption{Schematic of the quantum erasure: the double-slit in the
plane $x_{o}$ is imaged in the plane $x_{I}$ because of the
quantum correlations of entangled photon pairs. Hence the
which-path information is mapped onto the two-photon
imaging plane.}  \label{Fig1}       
\end{figure*}

The principle behind our realization of quantum erasure is
illustrated in Fig.~\ref{Fig1}. The entangled signal and idler
photons generated from SPDC are separated and directed to two
photon counting detectors through two individual arms of an
optical setup.  In one arm the signal photon passes through a
standard Young's double-slit interferometer; in the other arm, an
imaging lens is used for the production of the equal size
two-photon ``ghost" image  of the double-slit. There is an exact
point-to-point correspondence between the plane of the slits
$x_{o}$ and the image plane $x_{I}$ hence the information about
the path of the signal photon in the double-slit plane is mapped
onto the idler beam in the two-photon imaging plane.  At this
point we can choose to \textit{erase}  or to \textit{read} such
information to decide if the wave-like behavior, i.e., the
interference pattern, of the photon is observable.  To achieve
this, a Fourier transform approach is employed as shown in
Fig.~\ref{Fig2}.   The two-photon image function $f(x_{I})$, that
contains the which-path information, is Fourier transformed by the
lens $L'$ onto its Fourier transform plane.  On the Fourier
transform plane, the photon counting detector $D_{2}$ either
\emph{reads} the full transformed function or \emph{erases} most
of it. Knowledge of all the coefficients of the Fourier expansion
is sufficient to reconstruct the two-photon image function of the
double-slit that means knowing the which-path information.  On the
other hand, if only the DC term of the Fourier expansion is read,
it will never be possible to reconstruct the structure of the
image function $f(x_{I})$. Consequently, the which-path
information of the signal photon is \emph{erased}.  Thus, the wave
behavior will be learned by the observation of the interference.

The Fourier Transform approach to the quantum eraser is very
interesting, it provides more flexibility in the scheme. Since we
are transferring the which-path information through an image, one
could avoid using a double-slit that only provides two possible
paths. In principle an infinite number of paths or any spatial
information can be transferred and subsequently read or erased.

\section{Theory}
In this section we will first show that the which-path information
is indeed present in the two-photon imaging plane following the
exemplar setup in Fig.~\ref{Fig1} and then we will explain in
detail the two ways of collecting the idler photons that, for the
sake of clarity, we named \textit{erasing} and \textit{reading}
conditions (see Fig.~\ref{Fig2}).

\subsection{Mapping the which-path information in the ``ghost"
imaging plane}

In quantum theory of photodetection, the probability of having a
joint photodetection at two space-time points, $(\vec{r}_{1},
t_{1})$ and $(\vec{r}_{2}, t_{2})$, is governed by the second
order Glauber correlation function \cite{Glauber}:
\begin{eqnarray}\label{G2}
& & G^{(2)}(t_{1},\vec{r}_{1}; t_{2},\vec{r}_{2})  \equiv \\
\nonumber \langle & &
E_{1}^{(-)}(t_{1},\vec{r}_{1})E_{2}^{(-)}(t_{2},\vec{r}_{2})
E_{2}^{(+)}(t_{2},\vec{r}_{2})E_{1}^{(+)}(t_{1},\vec{r}_{1})\rangle.
\end{eqnarray}
where $E^{(-)}$ and $E^{(+)}$ are the negative-frequency and the
positive-frequency field operators at space-time points
$(\vec{r}_{1}, t_{1})$ and $(\vec{r}_{2}, t_{2})$ and the average
is done over the state of the radiation. Ignoring the temporal
part, the transverse electric field can be written as:
\begin{eqnarray}\label{E12}
\vec{E}^{(+)}_{1}(\vec{x}_{1})\propto\sum_{\vec{q}}g_{1}(\vec{x}_{1};\vec{q})\hat{a}(\vec{q})\\
\nonumber
\vec{E}^{(+)}_{2}(\vec{x}_{2})\propto\sum_{\vec{q}}g_{2}(\vec{x}_{2};\vec{q})\hat{a}(\vec{q})
\end{eqnarray}
where $\vec{x}_{i}$ is the transverse position of the $i^{th}$
detector, $\vec{q}$ is the transverse component of the momentum,
$\hat{a}(\vec{q})$ is the annihilation operator for the mode
corresponding to $\vec{q}$ and $g_{i}(\vec{x}_{i};\vec{q})$ is the
Green's function associated to the propagation of the field from
the source to the $i^{th}$ detector.

As far as the radiation is concerned, the process of SPDC involves
sending a pump laser beam into a nonlinear material. Occasionally,
the nonlinear interaction leads to the annihilation of a high
frequency pump photon and the creation of two lower frequency
photons known as signal and idler that satisfy the phase-matching
conditions \cite{klyshkobook,rubin}. The transverse part of the
state of the signal-idler radiation produced by a CW laser can be
simplified as follows:
\begin{eqnarray}\label{state}
 |\psi\rangle \propto\sum_{\vec{q},\vec{q}'}\delta(\vec{q}+\vec{q}')\hat{a}^{\dag}(\vec{q})\hat{a}^{\dag}(\vec{q}')|0\rangle.
\end{eqnarray}
In this case the second order correlation function can be written
as
\begin{equation}\label{g2spdc}
  G^{(2)}(\vec{r}_{1},t_{1};\vec{r}_{2},t_{2})=|\langle
0|E^{(+)}(\vec{r}_{2},t_{2})E^{(+)}(\vec{r}_{1},t_{1})|\psi\rangle|^{2}
\end{equation}
where $\langle 0|$ denotes the vacuum state and $|\psi\rangle$ the
two photon state of SPDC.
$\langle0|E^{(+)}(\vec{r}_{2},t_{2})E^{(+)}(\vec{r}_{1},t_{1})|\psi\rangle$
is an effective two-photon wavefunction, often referred to as
\textit{biphoton}.

While in classical optics intensities are measured, in two-photon
optics rate of joint detection counts, hence second order
correlation functions, are measured. And while in classical optics
intensities are the modulo-squared of electric fields, in
two-photon optics second order correlation functions are the
modulo-squared of the two-photon effective wavefunction. The
two-photon effective wavefunction contains the coherent
superposition of all the two-photon probability amplitudes that
can lead to a joint photodetection. This is the link between
classical Fourier optics and two-photon Fourier optics: the
results are equivalent if the classical electric field are
replaced by the two-photon probability amplitudes.

By using Eq.~\ref{E12} and Eq.~\ref{state}, the spatial part of
the second order correlation function reduces to:
\begin{eqnarray}\label{g2imaspdc}
 G^{(2)}(\vec{x}_{1},\vec{x}_{2})\propto|\sum_{\vec{q}}g_{1}(\vec{x}_{1},\vec{q})g_{2}(\vec{x}_{2},-\vec{q})|^{2}
\end{eqnarray}
where $\vec{x}_{1}$ and $\vec{x}_{2}$ are two-dimensional vectors
in the transverse planes of detectors $D_{1}$ and $D_{2}$
respectively.

Let's consider, first, the setup in Fig.~\ref{Fig1} in order to
show how the which-slit information is mapped into the ``ghost"
imaging plane. For the sake of simplicity, let's work in one
dimension just analyzing the horizontal transverse direction. For
this setup, the Green's functions are:
\begin{eqnarray}\label{g12}
g_{1}(x_{1};q)&\propto& \Psi[q,-\frac{c}{\omega}d_{A}]\int dx_{o}
T(x_{o})e^{i q x_{o}} \Psi[x_{1},\frac{\omega}{c} d_{A}']e^{i \frac{\omega x_{1}x_{o}}{c d_{A}'}}\nonumber\\
g_{2}(x_{2};q)&\propto&
\Psi[q,-\frac{c}{\omega}(d_{B}-\frac{1}{\frac{1}{d_{B}'}-\frac{1}{f}})]e^{\frac{i
q
x_{2}}{1-d_{B}'/f}}\Psi[x_{2},\frac{c}{\omega}\frac{1}{d_{B}'-f}]
\end{eqnarray}
where the paraxial approximation and a source of infinite
transverse size have been assumed.
$\Psi(|\vec{q}|,\frac{\omega}{c}p)=e^{\frac{i}{2}\frac{\omega}{c}p
|\vec{q}|^{2}}$ \cite{rubin}.

If the two-photon Gaussian thin lens equation is satisfied:
\begin{equation}\label{gauss}
\frac{1}{d_{A}+d_{B}}+\frac{1}{d_{B}'}=\frac{1}{f}
\end{equation}
and in particular the unitary magnification condition:
\begin{eqnarray}\label{m}
d_{A}+d_{B}& = & 2f \nonumber\\
d_{B}'& = & 2f
\end{eqnarray}
the second order correlation function can be rewritten as
\begin{eqnarray}\label{imagspdc}
 G^{(2)}(x_{1},x_{2})& \propto & |\int dx_{o}T(x_{o})e^{i \frac{\omega}{c d_{A}'}
 x_{1}x_{o}}\delta(x_{o}-x_{2})|^{2}\nonumber \\ & \propto &
 |T(x_{2})|^{2}
\end{eqnarray}

It is evident from the $\delta$-function in Eq.~\ref{imagspdc}
that every point of the plane of the double-slit is  linked to a
point in the ``ghost" imaging plane that we labelled $x_{I}$:
hence in the plane $x_{I}$ of the two-photon image there is the
information about the path followed by the signal photon in the
plane $x_{o}$ of the slit.

\subsection{Reading or erasing the which-path information of the ghost
imaging plane}

In order to have the ability of reading or erasing the which-path
information present at $x_{I}$ it is possible to place a second
lens $L'$ in the plane $x_{I}$ of the two-photon image and to
detect the idler photon in the plane where the Fourier transform
of the two-photon image is formed. It is known from Fourier
analysis applied to optical signals that if we measure all the
Fourier transform of the image we will still have all the
information we had in the image plane, but if we detect only one
point of such Fourier transform plane, the information of the
image plane will be inevitably erased.

\begin{figure*}
\resizebox{0.75\textwidth}{!}{%
  \includegraphics{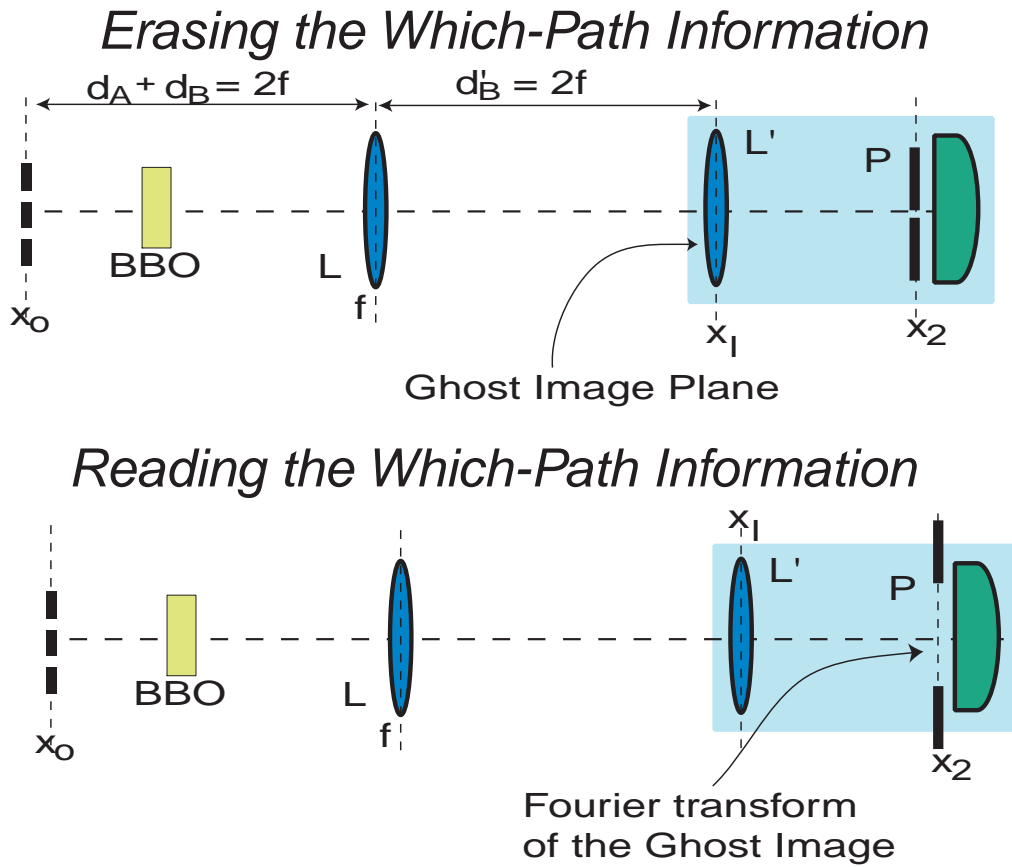}
}
 \caption{Klyshko's picture of the two-photon imaging setup showing
the two choices that we named \textit{erasing} and
\textit{reading}. In both cases a lens $L$ is placed in the plane
of the two-photon image and the detector is placed in the plane of
the Fourier transform of the two-photon image. In part (a) the
entire Fourier transform is collected by $D_{2}$.  In part (b)
only the central part of the Fourier transform is detected by
$D_{2}$.} \label{Fig2}       
\end{figure*}

The optical setup that implements such situation is depicted in
Fig.~\ref{Fig2}: a lens $L'$ of focal length $f'$ is in the
imaging plane and a pinhole $P$ is located before detector $D_{2}$
in the Fourier transform plane of the image field distribution. In
this case, the Green's function of one arm $g_{1}(x_{1};q)$ is
unchanged, while $g_{2}(x_{2};q)$ becomes (notice that $x_{2}$ is
in the plane of the detector $D_{2}$, and we will use $x_{I}$ to
indicate the transverse coordinate in the plane of the image):

\begin{eqnarray}\label{g2new}
g_{2}(x_{2};q)& \propto &
\Psi[q,-\frac{c}{\omega}(d_{B}+\frac{d_{B}' f}{f-d_{B}'})] \nonumber\\
\int d& x_{I} & e^{\frac{i q
x_{I}}{1-d_{B}'/f}}\Psi[x_{2},\frac{\omega}{c z}]e^{\frac{i \omega
x_{I}}{c z}}
\end{eqnarray}

The second order correlation function is then:

\begin{eqnarray}\label{fou1}
 G^{(2)}(x_{1},x_{2})& \propto &
 |\int dx_{I}T(x_{I})e^{i\frac{\omega x_{I}}{c}[\frac{x_{2}}{z}+\frac{x_{1}}{d_{A}'} ]}|^{2}
\end{eqnarray}

In the case in which only one point (e.g. $x_{2}=0$) in the
Fourier plane is considered (what we named \textit{erasing}
condition), the second order correlation function reads:

\begin{eqnarray}\label{foures}
 G^{(2)}_{erase}(x_{1})& \propto &
 |\int dx_{I}T(x_{I})e^{i \frac{c x_{1}x_{I}}{\omega
 d_{A}'}}|^{2}\nonumber \\
 & \propto & |\textit{F}_{\frac{\omega x_{1}}{c d_{A}'}}(T(x_{I}))|^{2}
\end{eqnarray}
that in the case of a double-slit of slit width $a$ and slit
separation $d$ becomes the usual interference diffraction pattern:

\begin{eqnarray}\label{slits}
 G^{(2)}_{erase}(x_{1})& \propto & Sinc^{2}(\frac{\pi x_{1} a}{\lambda d_{A}'})Cos^{2}(\frac{\pi x_{1} d}{\lambda d_{A}'})
\end{eqnarray}

In the case in which all the photons arriving in the Fourier plane
are detected (\textit{reading} condition), the second order
correlation function becomes:

\begin{eqnarray}\label{totinc}
G^{(2)}_{read}(x_{1})\propto \int dx_{2}
|\textit{F}_{\frac{\omega}{c}[\frac{x_{1}}{d_{A}'}+\frac{x_{2}}{z}]}\{T(x_{I})\}|^{2}
= constant
\end{eqnarray}
that shows the absence of any interference pattern.

As we pointed out in the introduction, the interpretation of the
quantum eraser results in terms of complementarity examination on
a single photon is troubling. On the other hand, the
straightforward calculation presented here can be intuitively
captured if it is based on the concept of nonlocal two-photon
amplitudes and their coherent superposition. In this sense, the
physics behind entangled two-photon phenomena seems having no
classical counterpart in electromagnetic theory. In order to help
clarifying this physics, Klyshko proposed an ``advanced-wave
model"\cite{klyshko} that forces a classical counterpart of the
concept of two-photon amplitudes and the associated two-photon
optics. In his model, Klyshko considered the light to start from
one of the detectors, propagate backwards in time until the
two-photon source of SPDC and then forward in time towards the
other detector.  The two-photon source is thus playing the role of
a mirror to keep the proper transverse momentum relation of the
entangled photon pair. Fig.~\ref{Fig2} is particularly suitable
for Klyshko's picture: the which path information is carried by
the advanced waves from the double-slit to the Fourier Transform
plane of the ghost image according to the classical rules of
Fourier optics.  A straightforward calculation reveals that the
Fourier Transform plane, referring to Fig.~\ref{Fig2}, is at a
distance $z$ from lens $L'$ such that:

\begin{equation}\label{imagefourier}
\frac{1}{d_{B}'-f}+ \frac{1}{z}=\frac{1}{f'}
\end{equation}

This equation has a ready explanation: it is a thin lens equation
involving lens $L'$ in which the object plane coincides with the
focal plane of lens $L$. Therefore, if we read Fig.~\ref{Fig2}
from right to left, from the idler detector till the double-slit,
we can provide another perspective on the optical interpretation
of the phenomenon. The erasing condition is equivalent to having a
point source and a lens system at a focal distance from it in such
a way that only one momentum of propagation of the two-photon
light is selected. It is then natural to observe an interference
pattern from the double-slit because collimated radiation, with
only one $\vec{k}$, is impinging on it. On the other hand, in the
reading condition the situation is equivalent to having an
extended source; as a result the radiation that impinges on the
double-slit has all possible values of the momentum and each of
the momenta will produce a slightly shifted
interference-diffraction pattern. The total result, due to the
incoherent sum of all such patterns, will be a constant.

The above two-photon picture helps establishing a connection
between this eraser and the ghost interference effect first
demonstrated by Strekalov et al. \cite{strekalov} as well as the
erasure's idea by Dopfer et al. that similarly used the transverse
correlations of SPDC even though it did not involve the transfer
of the which-path information via imaging\cite{dopfer}. Using
Klyshko's picture, even the puzzling physics of quantum erasure is
trivial, which is the beauty of Klyshko's model.

\section{Experiment}
A sketch of the experimental setup is presented in
Fig.~\ref{Fig3}. A $5$-mm type-II BBO crystal, cut for collinear
degenerate phase matching was pumped by an Ar$^{+}$ laser at
wavelength $457.9$ nm. After passing the nonlinear crystal, the
pump radiation was filtered out by a mirror with high reflection
at the pump wavelength and high transmission at the wavelength of
the signal and idler by an $RG715$ color glass filter. The
signal-idler radiation was then split by a polarizing beam
splitter; in the transmitted arm ($A$) a double-slit was placed at
distance $d_{A}=115 mm$ from the crystal and in the far field zone
($d_{A}'=1250 mm$) a narrow bandpass filter ($10$ nm band centered
at $916$ nm) was inserted  in front of $D_{1}$, a single photon
counting detector (Perkin Elmer SPCM-AQR-$14$) that was used to
scan the transverse horizontal direction; in the reflected arm
($B$) a lens $L$ of focal length $f=500 mm$ was placed at a
distance $d_{B}=885 mm$ from the BBO crystal and a non polarizing
beam splitter ($NPBS$) was at a distance $d_{NPBS}=985 mm$ from
the lens $L$. Notice that the beam splitter $NPBS$ is the device
at which the idler photon makes the random choice. In the output
ports of $NPBS$ we built the two different ways of detecting the
idler photons following the example of Fig.~\ref{Fig4}: in the
transmitted arm a lens $L_{T}'$ of focal length $f_{T}'=250 mm$
was at a distance $d_{L'}=15 mm$ from $NPBS$ and a very small
pinhole $P_{T}$ was placed at $z_{T}=500 mm$ from the lens just
before coupling the radiation in a $4.5$ m long multimode optical
fiber ($F_{T}$); in the reflected arm a lens $L_{R}'$ of focal
length $f_{T}'=50 mm$ was at a distance $d_{L'}=15 mm$ from $NPBS$
and a completely open pinhole $P_{R}$ was placed at $z_{R}=55 mm$
from the lens just before coupling the radiation in a $2$ m long
multimode optical fiber ($F_{R}$). The two optical fibers were
then joined at a $2$ to $1$ fiber combiner and their output was
filtered by a narrow bandpass filter ($10$ nm band centered at
$916$ nm) and measured by a single photon counting detector
(Perkin Elmer SPCM-AQR-$14$). The output photocurrent pulses from
the two photodetectors were finally sent to the ``start" and
``stop" inputs of a Time to Amplitude Converter (TAC) then
connected to a MultiChannel Analyzer (MCA) and with a PC the
coincidence counting rate in a desired window was measured.

Usually in delayed-choice quantum eraser experiments, each choice
is associated with a different detector and therefore the two
situations of no interference or recovered interference are
obtained by counting coincidences with a separate measurement
device. In our experiment we decided to use only one
photo-detector $D_{2}$. Both transmitted and reflected photons at
$NPBS$ were sent to $D_2$ with different optical delays (given by
the different length of the fibers $F_{T}$ and $F_{R}$). In this
way we created histograms as in Fig.~\ref{Fig4} that measure the
second-order correlation function as a function of $t_{2}-t_{1}$
and that calibrate the coincidence time window for the actual
coincidence counting measurement: the first peak corresponds to
the coincidence counts of $D_{2}$ with the reflected side of
$NPBS$ (\textit{reading} situation) while the second peak
corresponds to the coincidence detections of $D_{2}$ with the
transmitted side of $NPBS$ (\textit{erasing} condition). The
coincidence counting rate associated to each choice is then
measured within the appropriate coincidence time window.

The curves in Fig.~\ref{Fig4} carry another significant
information. The FWHM of the curves is mainly determined by the
response times of the detectors and measures the uncertainty with
which we are able to determine the difference in time arrival
between signal and idler photons $\Delta (t_{2}-t_{1})$. In order
to achieve the delayed erasure condition, i.e. the choice of the
idler photon and the detection of the signal photon at $D_{1}$
have to be space-like separated events; the optical path
difference between the crystal and the $NPBS$ (where the choice is
randomly made) has to be bigger that the distance from the crystal
to detector $D_{1}$ of a quantity larger than $\Delta
(t_{2}-t_{1})$. In this case $\Delta (t_{2}-t_{1}) \sim 1 ns$,
while the difference in path lengths
$[(d_{B}+d_{NPBS})-(d'_{A}+d_{A})]/c \sim 1.7 ns$, therefore we
can be sure that the choice is made after the detection of the
signal photon at detector $D_{1}$.

The experimental results are shown in Fig.~\ref{Fig5} and
Fig.~\ref{Fig6}. They refer to two different double-slits: one
with slit width of $a=150 \mu m$ and slit separation $d=470 \mu m$
and the other with slit width of $a=100 \mu m$ and slit separation
$d=250 \mu m$. In the graphs both the \textit{erasing} condition
measurement (empty circles) and the \textit{reading} condition
measurements (filled squared) are shown. As expected, when we read
the which-path information, we do not see any interference pattern
while when we erase such information the experimental data agree
with the expected interference-diffraction of the double-slits.
The visibilities of both interference patterns are very high ($85
\%$ and $95 \%$ respectively) and only limited by the finite size
($\sim 200 \mu m$) of detector $D_{1}$.

Let us now point out some additional characteristics of the
experimental setup. ($1$) The double-slit has to satisfy the
condition $\Delta \theta \gg \lambda/d$ in order to avoid the
existence of any first order interference-diffraction pattern.
From the tuning curves of the BBO crystal we computed the
divergence of the SPDC radiation to be around $\Delta \theta \sim
27 mrad$. For this reason we chose two different double-slits, one
with $\lambda /d \sim 3 mrad$ and the other with $\lambda/d \sim
1.5 mrad$. ($2$) The two lenses $L_{T}'$ and $L_{R}'$ are placed
in the plane where the two-photon image of the double-slit is
formed as described in the theory section. Notice, in fact, that
the distance from the slit, back to the crystal and forward to the
lens $L$, i.e. $d_{A}+d_{B}$ is exactly equal to $2f$; also, the
distances from the lens $L$ to the two lenses $L_{T}'$ and
$L_{R}'$, i.e. $d_{B}'=d_{NPBS}+d_{L'}$ is again equal to $2f$.
Therefore such distances satisfy the two-photon Gaussian thin lens
equation with unitary magnification of Eq.~\ref{gauss} and
Eq.~\ref{m}. ($3$) We used two different lenses $L'$ in order to
better achieve the \textit{reading} and \textit{erasing} condition
mentioned in the theory section. In the transmitted arm of $NPBS$,
the focal length of $L_{T}'$ is large, therefore a small pinhole
is a good approximation of taking only the central point of the
Fourier plane; on the other hand, in the reflected arm the focal
length of $L_{R}'$ is short, therefore a completely open pinhole
$P_{R}$ of diameter $\sim 1cm$, is a very good approximation of
detection of all the Fourier plane.

\section{Conclusion}
The key idea of the eraser is to transfer the which-path
information to a distant location via a two-photon ``ghost" image
and then \textit{read} or \textit{erase} the path information in
its Fourier transform. Therefore, the result of this quantum
eraser can be viewed in terms of continuous variables of position
and momentum. This aspect is interesting given the recent interest
in continuous variable entanglement for quantum information
processing\cite{leuchs,filip,marek}. Using a double slit in the
actual experiment and therefore proving Eq.~\ref{slits} was a
matter of convenience and clarity. However, the observed
interference-diffraction pattern is experimental evidence of the
general result in Eq.~\ref{foures}.  Therefore, there is no
restriction, in principle, to extend the present idea to
multi-slits (paths) or even continuous spatial modulation.

Having demonstrated a quantum eraser via two-photon ghost imaging
with a Fourier transform approach shows that ghost imaging schemes
coherently transfer the optical information between two distinct
arms of a setup. This property might be useful because it shows
the possibility to implement phase operations, or Fourier
manipulations in a nonlocal fashion to improve the optical
performances of imaging schemes.

From a fundamental point of view, we have demonstrated a new
scheme for delayed choice quantum eraser.  This new eraser has
probed all the interesting physics proposed by Scully and Druhl.
The experiment, from a different perspective, demonstrates and
questions  two of the most intriguing fundamental concepts of
quantum theory: complementarity and entanglement.

As for the complementarity, for the first time, a delayed choice
quantum eraser is demonstrated in which the choice to erase or not
erase is realized truly at random in only one photo-detector. In
many previous experiments the interference or no interference
situations involved basically different experiments and/or
different experimental runs. Some of the other previous quantum
erasers did not involve different experimental realizations, but
anyway two different measurement devices were associated to the
erasing and not erasing conditions. In our experiment all the
photons, belonging to either choice, arrive to the same
photo-electric device truly at random. This aspect is very
interesting because the experimental conditions associated with
the different choices are very similar and therefore it is not
trivial to establish where the measurement, hence the collapse of
the wave function, occurs.

As for the entanglement, this experiment has strikingly shown a
fundamental point that is often forgotten: for entangled photons
it is misleading and incorrect to interpret the physical phenomena
in terms of independent photons. On the contrary the concept of
``biphoton" wavepacket has to be introduced to understand the
nonlocal spatio-temporal correlations of such kind of states.
Based on such a concept, a complete equivalence between two-photon
Fourier optics and classical Fourier optics can be established if
the classical electric field is replaced with the two-photon
probability amplitude.  The physical interpretation of the eraser
that is so puzzling in terms of individual photons' behavior is
seen as a straightforward application of two-photon imaging
systems if the nonlocal character of the biphoton is taken into
account by using Klyshko's picture.

The authors thank M.H. Rubin for everyday help and discussions and
M.O. Scully and C.O. Alley for the encouragement of conducting
this experiment. This research was supported in part by ARO and
the CASPR program of NASA.

%

%

\begin{figure*}
\resizebox{0.75\textwidth}{!}{
  \includegraphics{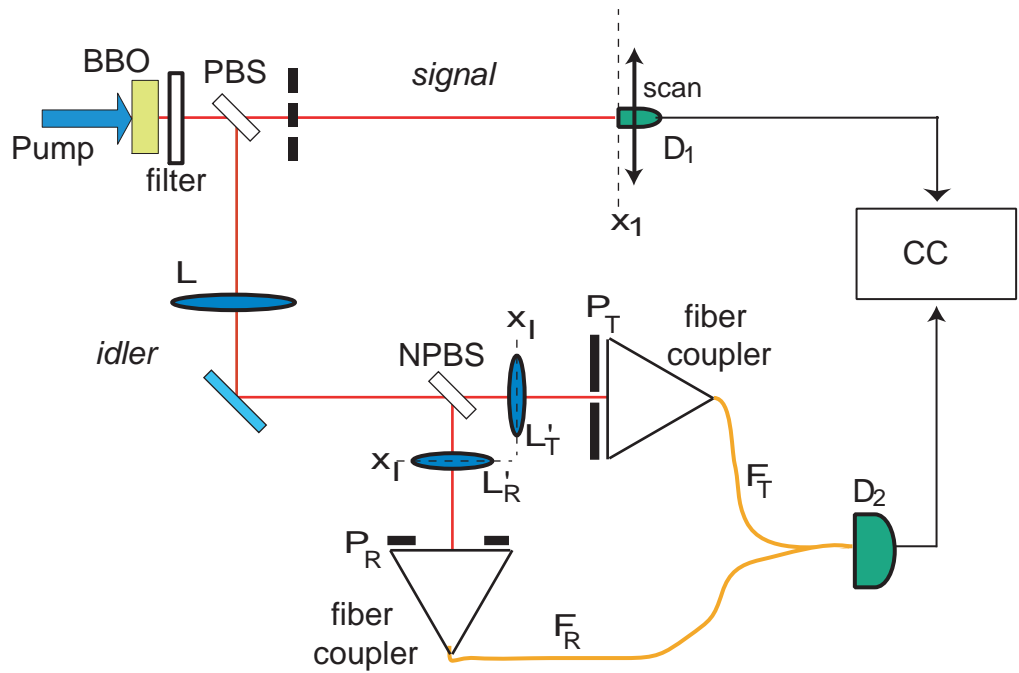}
}
 \caption{Schematic of the experimental setup.} \label{Fig3s}       
\end{figure*}

\begin{figure*}
\resizebox{0.75\textwidth}{!}{
  \includegraphics{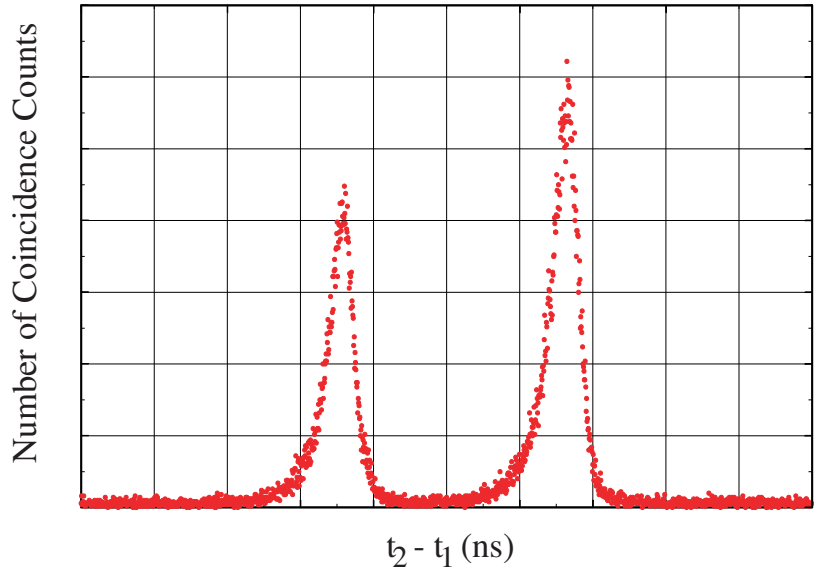}
}
 \caption{Typical MCA distribution. The two different way of
detecting idler photon are time ``encoded" by using two multimode
optical fibers of different length; hence in the MCA pattern the
two peaks correspond to the two different situations.} \label{Fig4}       
\end{figure*}

\begin{figure*}
\resizebox{0.75\textwidth}{!}{
  \includegraphics{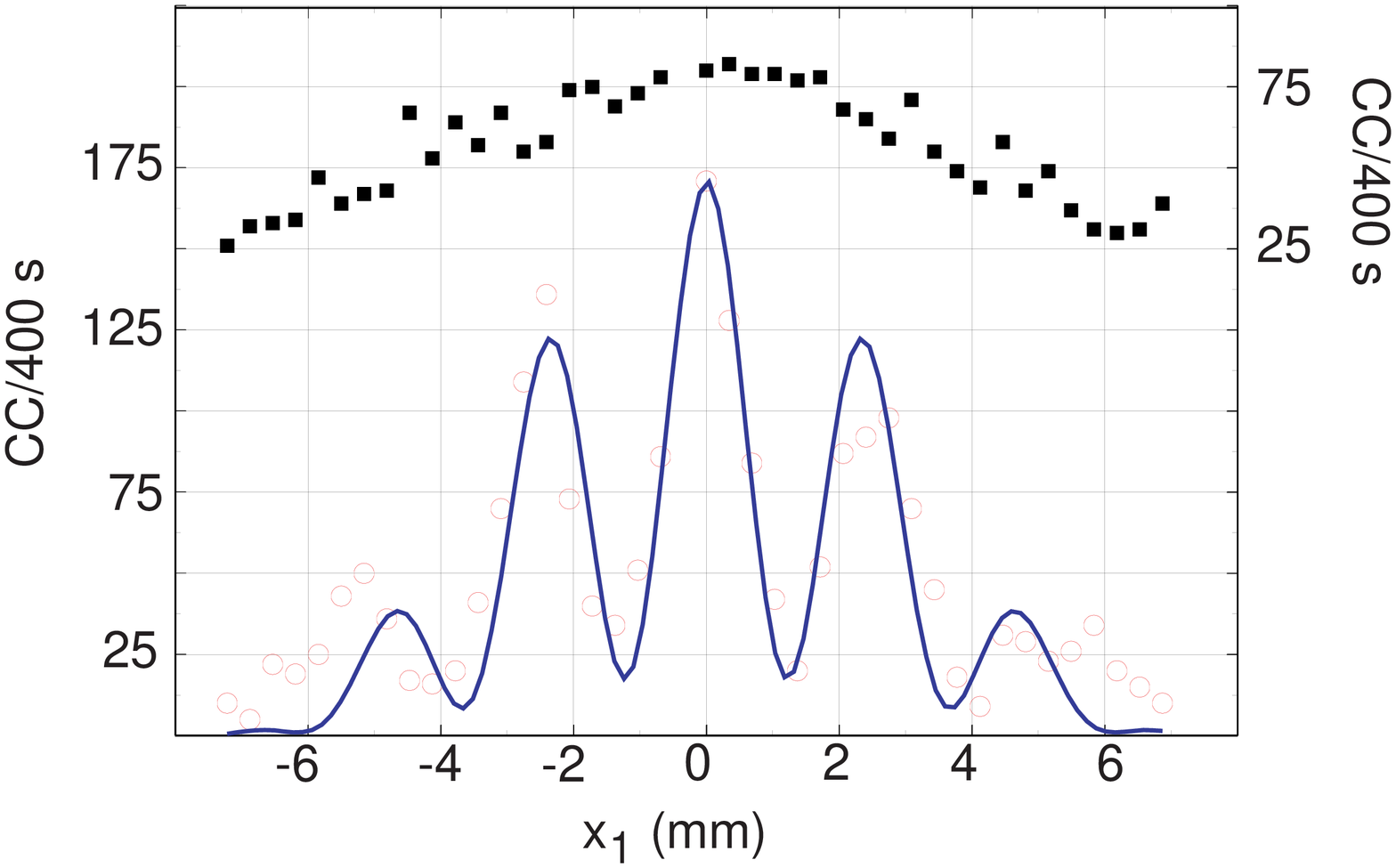}
}
 \caption{Experimental results for a double-slit of slit width
$a=150 \mu m$ and slit separation $d=470 \mu m$. The filled
squares show the coincidence count pattern obtained in the
\textit{reading} situation, while the empty circles indicates the
pattern obtained in the \textit{erasing} situation. The solid line
is the theoretical expectation of a $85 \%$ visibility
interference-diffraction pattern.} \label{Fig5}       
\end{figure*}

\begin{figure*}
\resizebox{0.75\textwidth}{!}{
  \includegraphics{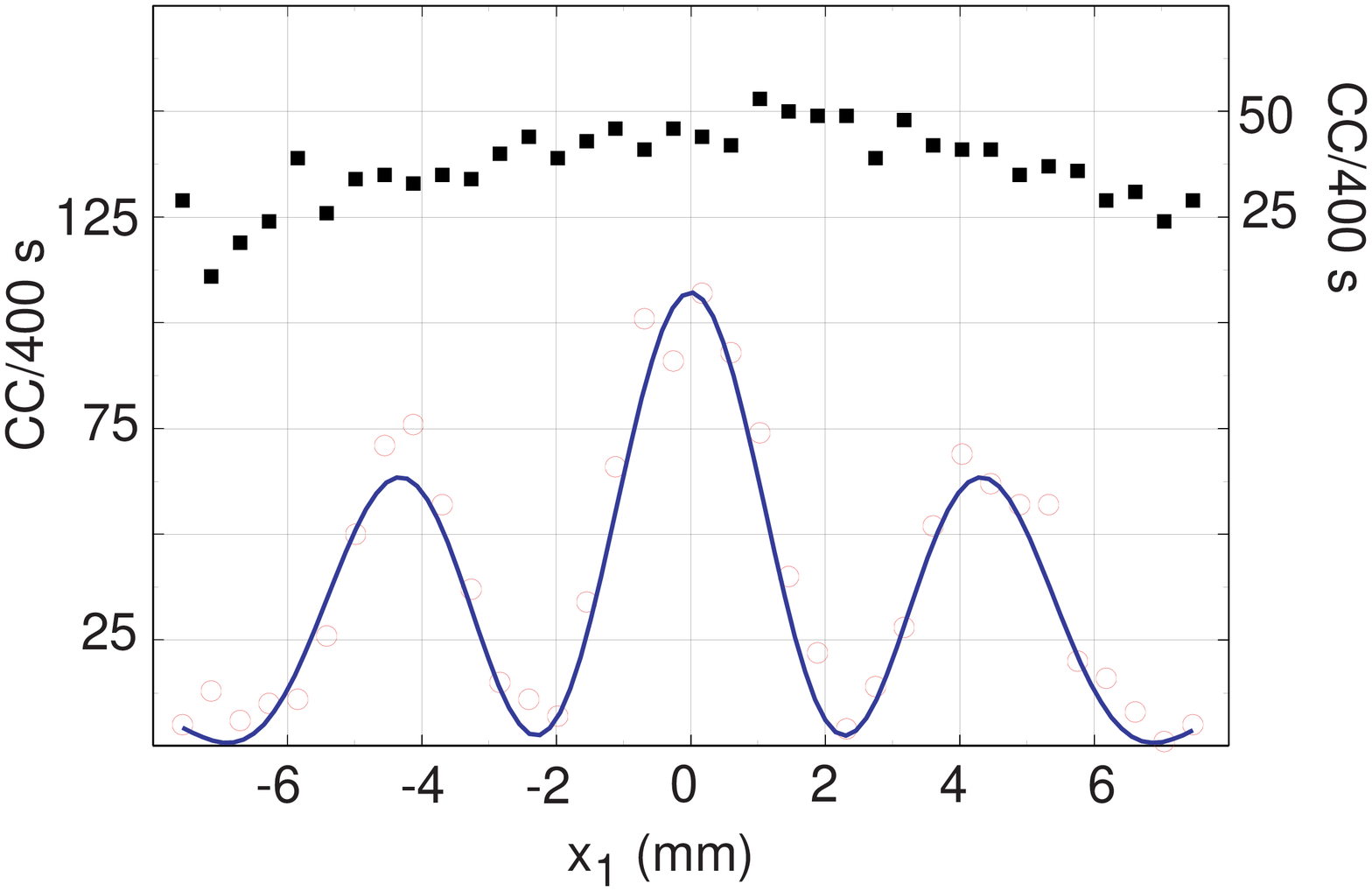}
}
 \caption{Experimental results for a double-slit of slit width
$a=100 \mu m$ and slit separation $d=25 \mu m$. The filled squares
show the coincidence count pattern obtained in the
\textit{reading} situation, while the empty circles indicates tha
pattern obtained in the \textit{erasing} situation. The solid line
is the theoretical expectation of a $95 \%$ visibility
interference-diffraction pattern.} \label{Fig6}       
\end{figure*}

\end{document}